\documentclass[aps,twocolumn,prl,showpacs,amsmath]{revtex4}
\usepackage{graphicx}
\begin{document}

\title{Evidence for Superfluidity in a Resonantly Interacting Fermi Gas}
\author{J. Kinast}
\author{S. L. Hemmer}
\author{M. E. Gehm}
\author{A. Turlapov}
\author{J. E. Thomas}
\affiliation{Physics Department, Duke University, Durham, North Carolina 27708-0305}

\date{\today}

\begin{abstract}
We observe collective oscillations  of a trapped, degenerate Fermi
gas of $^6$Li atoms at a magnetic field just above a Feshbach
resonance, where the two-body physics does not support a bound
state.  The gas exhibits a radial breathing mode at a frequency of
2837(05) Hz, in excellent agreement with the frequency of
$\nu_H\equiv\sqrt{10\nu_x\nu_y/3}=2830(20)$ Hz predicted for a
{\em hydrodynamic} Fermi gas with unitarity limited interactions.
The measured damping times and frequencies are inconsistent with
predictions for both the collisionless mean field regime and for
collisional hydrodynamics. These observations provide the first
evidence for superfluid hydrodynamics in a resonantly interacting
Fermi gas.
\end{abstract}

\maketitle

Strongly-interacting two-component Fermi gases provide a unique
testing ground for the theories of exotic systems in nature,
ranging from super-high temperature superconductors to neutron
stars and nuclear matter. The feature which all of these systems
have in common is a strong interaction between pairs of spin-up
and spin-down particles. In atomic Fermi gases, tunable, strong
interactions are produced using a Feshbach
resonance~\cite{Houbiers,Stwalley,Tiesinga}. Near the resonance,
the zero-energy s-wave scattering length $a$ exceeds the
interparticle spacing, and the interparticle interactions are
unitarity-limited and universal~\cite{Heiselberg,MechStab,Ho}. In
this region, high temperature Cooper pairing has been
predicted~\cite{Stoof1,Holland,Timmermans,Griffin}.

In this Letter, we present measurements of the frequencies and
damping times for the radial hydrodynamic breathing mode of a
trapped, highly degenerate gas of $^6$Li atoms just above a
Feshbach resonance~\cite{GrimmThomas}. These observations  provide
the first evidence of superfluid hydrodynamics in a resonantly
interacting two-component Fermi gas.

Hydrodynamic behavior in a collisionless quantum gas at very low
temperature is known to be a hallmark of superfluidity.
Previously, we observed hydrodynamic, anisotropic expansion of a
strongly interacting, ultracold, two-component Fermi
gas~\cite{OHaraScience}. However, an initially collisionless gas
could have become collisionally hydrodynamic as the Fermi surface
significantly deformed during the expansion~\cite{Gupta}. Thus,
the observations suggested superfluid hydrodynamics, but were not
conclusive.

Superfluidity has been observed in Bose-Einstein condensates
(BECs) of molecular dimers, which have been produced from a
two-component strongly interacting Fermi gas. The first
experiments produced the dimer BECs at a magnetic field below the
Feshbach resonance, where the atoms have a nonzero binding
energy~\cite{GreinerBEC,JochimBEC,ZwierleinBEC,Salomon}. Although
the two-body physics does not support a bound state at magnetic
fields above the Feshbach resonance, for fermionic atoms, the
many-body physics does. Observations of BECs originating from such
dimers are consistent with the existence of preformed
pairs~\cite{RegalCooper,ZwierleinCooper}. These experiments
indirectly explore the microscopic structure, while our
experiments are complementary in that they directly measure the
macroscopic dynamics.

One method for distinguishing between a BEC and a superfluid Fermi
gas is to examine their collective hydrodynamic modes at low
temperature where the trapped gas is collisionless~\cite{Vichi}.
For a weakly repulsive BEC contained in a nearly cylindrically
symmetric trap, the radial breathing mode occurs at a frequency of
\begin{equation}
\nu_B=2\sqrt{\nu_x\nu_y},
 \label{eq:Bosefreq}
 \end{equation} where
$\nu_i$ are the harmonic oscillation frequencies in Hz of a
noninteracting gas in the $i^{th}$ direction of the trap. In
contrast to a weakly repulsive BEC, a superfluid Fermi gas in a
cigar-shaped trap is predicted to have a radial breathing mode at
the hydrodynamic frequency
\begin{equation}\nu_H=\sqrt{\frac{10}{3}\nu_x\nu_y}.
\label{eq:Fermifreq}
\end{equation}
 This result is obtained in the
unitarity limit, where the shift from the interparticle
interactions vanishes for a hydrodynamic
gas~\cite{Stringari,Heiselberg2}. In general, hydrodynamics with
the frequency $\nu_H$ can arise from superfluidity or from
collisions in a normal fluid. However, at low temperature, Pauli
blocking is expected to suppress the collision rate in a Fermi
gas, {\em increasing} the damping of the collisionally
hydrodynamic modes as the temperature is lowered.

In our experiments with a trapped Fermi gas,  Pauli-blocking of
collisions is expected to be effective at the lowest temperatures
achieved. Nevertheless, a weakly damped radial mode at precisely
the hydrodynamic frequency of $\nu_H$ is observed, and the damping
rate {\em decreases} strongly as the temperature is lowered below
$\simeq 30$\% of the Fermi temperature for a noninteracting gas.
These observations provide the first evidence for superfluid
hydrodynamics in a resonantly interacting Fermi gas.

We measure the frequencies and damping times of the radial
breathing mode of the $^6$Li Fermi gas near the Feshbach resonance
at 822 (3) G~\cite{ZwierleinCooper}. The frequencies measured at
several temperatures and magnetic fields are compared to the
predictions (Eqs.~\ref{eq:Bosefreq} and~\ref{eq:Fermifreq}) based
on the trap oscillation frequencies measured by parametric
resonance. A cross check of the measurement method is provided by
measuring the breathing mode of the noninteracting gas.

We begin by preparing a degenerate, 50-50 mixture of the lowest
spin-up and spin-down states of $^6$Li atoms in an ultrastable
CO$_2$ laser trap as described previously~\cite{OHaraScience}.
Forced evaporation at a chosen magnetic field in the range 770-910
G produces a highly degenerate, unitarity-limited sample. The trap
depth is lowered by a factor of $\simeq 580$ over 4 s, then
recompressed to 4.6\% of the full trap depth  in 1 s  and held for
1 s to assure equilibrium. This produces a degenerate sample of
$N/2\simeq 1.5\times 10^5$ atoms per spin-state at a temperature
of $\simeq 0.15 \,T_F$. The corresponding Fermi temperature for a
noninteracting gas is then
$T_F=\hbar\bar{\omega}(3N)^{1/3}=2.5\,\mu$K
($\bar{\omega}=2\pi\times (\nu_x\nu_y\nu_z)^{1/3}$), small
compared to the final trap depth of $35\,\mu$K.

From the images of the released cloud, the number of atoms is
determined by integrating the column density. We have measured the
resolution of the entire imaging system to be $5.5\,\mu$m. For low
temperatures, $T/T_F \leq 0.4$, the ratio $T/T_F$ is determined by
fitting a Thomas-Fermi profile for a noninteracting Fermi gas to
the transverse (x) distribution obtained by integrating the column
density in the axial (z) direction. Very good fits are obtained
for the noninteracting gas at 526 G. At 910 G, this procedure
yields temperatures as low as $T/T_F=0.06$ and excellent fits.
However, at fields closer to resonance, slightly higher
temperatures are obtained, and the shape may not be precisely
Thomas-Fermi due to many-body effects. For measurements near
resonance at higher temperatures $T/T_F=0.5-1.2$, where the gas is
not perfectly hydrodynamic, temperature estimation is less
precise. Here, we simply assume hydrodynamic expansion of a
Maxwell-Boltzmann spatial distribution for a noninteracting gas.
The measured temperatures therefore indicate the trend but not
necessarily the absolute temperature. We find consistency between
the measured number of trapped atoms, the  temperature, and the
initial cloud size obtained by hydrodynamic
scaling~\cite{OHaraScience}. For the strongly interacting gas, we
include a reduction of the cloud radius arising from the mean
field~\cite{MechStab}. By correcting selected images for our
estimated saturation $I/I_{sat} = 0.2$, we estimate that the true
temperatures are lower by 0.03 $T_F$ and the true atom numbers are
increased by a factor near 1.15 compared to the values given in
Table~\ref{table:1}.

Trap oscillation frequencies at 4.6\% of the full well depth are
measured by parametric resonance in a weakly-interacting sample.
The gas is cooled by forced evaporation over 25 seconds to
temperatures of $0.3\,T_F$ at a field of 300 G and the trap depth
is then modulated by 0.5\% for 1 s. During this period, the low
collision rate produces little damping, but permits the gas to
thermalize. After modulation, imaging at 526 G is used to measure
the release energy versus drive frequency. Well-resolved
resonances are obtained at $2\omega_x=2\pi\times 3200(20)$Hz and
$2\omega_y=2\pi\times 3000(20)$ Hz. Because of the low frequency,
the axial resonance is measured at full trap depth. We obtain
$2\omega_z = 2\pi\times 600(20)$ Hz, yielding $2\omega_z =
2\pi\times 140(5)$ Hz  at 4.6\% of full trap depth, including a
quadratically combined magnetic field curvature contribution of 21
Hz. From these measurements,
$\nu_\perp\equiv\sqrt{\nu_x\nu_y}=1550(20)$ Hz.

To excite the transverse breathing mode, the trap is turned off
abruptly ($\leq 1\,\mu$s) and turned back on after a delay  of
$t_0=50\,\mu$s. To show that our excitation is a weak
perturbation, we estimate the energy increase, which arises
principally from the change in potential energy in the transverse
directions. Assuming approximately ballistic expansion, $\Delta
E_\perp=E_\perp(\omega_\perp t_0)^2/2=0.1\,E_\perp$. For initial
temperatures of 0.1-0.15 $T_F$, the corresponding temperature
change is $\Delta T/T_F\simeq 0.05$ when the gas thermalizes,
consistent with our measurements. To measure the frequency and
damping time, the breathing mode is excited and the sample is held
for a variable time $t_{hold}$, after which the trap  is
extinguished suddenly, releasing the gas. The cloud is  allowed to
expand for 1 ms and then imaged using a camera beam probe pulse of
5 $\mu$s duration arranged to produce a two-level system as
described previously~\cite{OHaraScience}.

\begin{figure}[h]
\begin{center}
\resizebox{3.5in}{!}{\includegraphics[1.3in,4.2 in][10in,7.25
in]{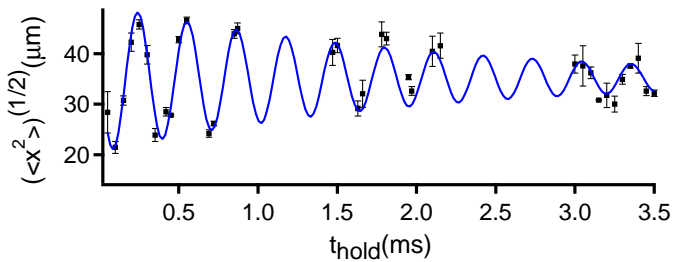}}
\end{center}
\caption{Excitation of the breathing mode in a noninteracting
Fermi gas of $^6$Li at 526 G. Error bars denote one-$\sigma$
uncertainty.} \label{fig:1}
\end{figure}

To study a noninteracting sample, breathing modes  are excited at
526 G, where the scattering length is nearly
zero~\cite{zerocross,Grimmzero}, after cooling at 300 G as
described above. Fig.~\ref{fig:1} shows $\sqrt{\langle
x^2\rangle}$ for the expanded gas ($\langle x\rangle\equiv 0$),
plotted as a function of the hold time in the trap, $t_{hold}$.
Fitting an exponentially damped sinusoid
$x_{\text{rms}}+A\exp(-t/\tau_{\text{damp}})\sin(2\pi \nu t
+\varphi )$ to the data, we obtain the frequency $\nu = 3212(30)$
Hz and the damping time  $\tau_{\text{damp}}=2.04$ ms, where the
errors are from the fit. The damping time is consistent with a
small anharmonicity of the gaussian profile of the trap potential.

\begin{figure}
\begin{center}
\resizebox{3.5in}{!}{\includegraphics[0.5in,2in][8.25in,11in]{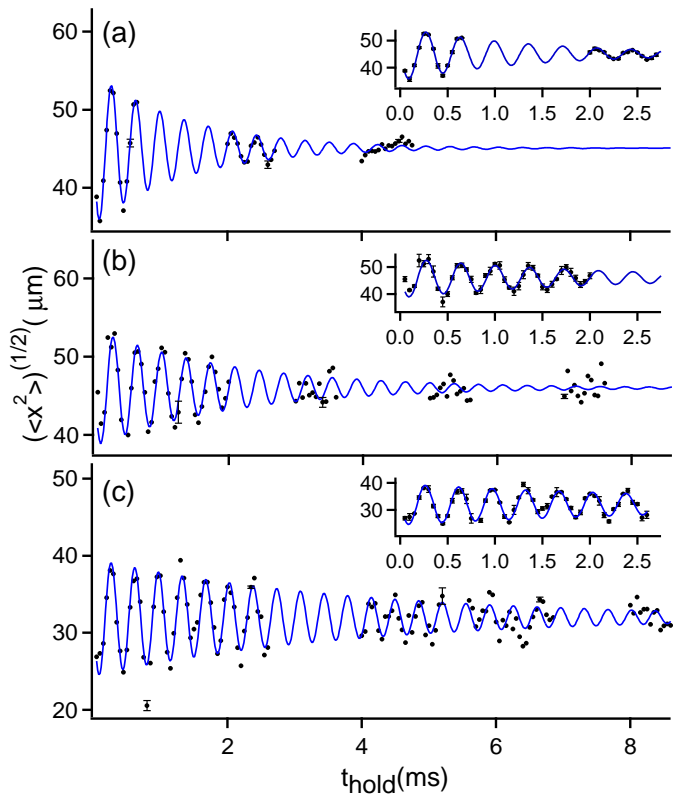}}
\end{center}
\caption{Excitation of the breathing mode in a strongly
interacting Fermi gas of $^6$Li at 870 G. a) $T/T_F=0.50$, b)
$T/T_F=0.33$ c) $T/T_F=0.17$. Error bars denote one-$\sigma$
uncertainty.} \label{fig:2}
\end{figure}

The strongly interacting gas exhibits longer damping times at low
temperature. Fig.~\ref{fig:2}(a-c) shows the results at 870 G as
the temperature is lowered from $T/T_F=0.50$ to 0.17. The damping
time increases from 1.4(0.1) ms to 3.85 (0.4) ms. At the lowest
temperature, the measured oscillation frequency  is $2837(05)$ Hz.

Results for different temperatures and magnetic fields are given
in Table~\ref{table:1}.

\begin{table}
\centering
\begin{tabular}{|c|c|c|c|c|c|}
  \hline
  B(G) & $T/T_F$ & N($10^3$)& $x_{\text{rms}}(\mu$m) & $\nu$(Hz) & $\tau_{damp}$(ms)  \\
  \hline
 526$^*$ & 0.30(.02)& 288(18)&35.2 & 3212(30) & 2.04(0.4)  \\ \hline
 770 & 0.13(.03)$^\dagger$ & 138(20)&29.3 &3000(150) & 2.00(1.1)  \\ \hline
  815 & 0.14(.04)$^\dagger$ & 198(24)& 24.0&2931(19) & 3.60(1.5)  \\ \hline
  860 & 0.14(.04)& 294(26) &28.6 & 2857(16) & 3.67(1.1)  \\ \hline
  870$^*$ & 0.17(.06) & 288(30) &32.0 & 2837(05) & 3.85(0.4)  \\ \hline
   870$^1$& 0.15(.03) & 225(36) &33.5 & 2838(06)& 6.01(1.4)\\ \hline
   870$^2$& 0.18(.04) & 207(28) &41.8 & 5938(18)& 1.44(0.2)\\ \hline
  870$^*$ & 0.33(.02)  & 379(24)&46.1 & 2754(14) & 2.01(0.3) \\ \hline
  870$^*$ & 0.50(.06) & 290(32)&45.1 & 2775(08) & 1.39(0.1)  \\ \hline

  870 & 1.15(0.10) & 244(10)&41.7 & 2779(50) &1.08(0.4)  \\ \hline
  880 & 0.12(.04) & 258(30) &30.0 & 2836(16) & 3.95(1.5) \\ \hline
  910 & 0.11(.06) & 268(17)&27.8 & 2798(15) & 3.30(1.1)  \\
  \hline
\end{tabular}
\caption{Breathing mode frequencies $\nu$ and damping times
$\tau_{damp}$. B is the applied magnetic field, $T/T_F$ is the
initial temperature in units of the Fermi temperature and N is the
total number of atoms, uncorrected for saturation.
$x_{\text{rms}}$ is the time-averaged root-mean-square size of the
oscillating cloud. $^\dagger$From the tails of a bimodal
distribution. $^*$Shown in the figures. $^1$For $t_0=25\,\mu$s.
$^2$At 18.8(0.9)\% trap depth, $t_0=25\,\mu$s and 0.8 ms expansion
time. Error estimates are from the fit only.}\label{table:1}
\end{table}

The measured oscillation frequencies for the transverse breathing
mode can be compared to those expected for a noninteracting gas,
for a weakly repulsive Bose gas, and for a hydrodynamic Fermi gas.
For a weakly repulsive Bose gas, using eqs.~\ref{eq:Bosefreq}
and~\ref{eq:Fermifreq}, we predict $\nu_B=2\nu_\perp =3100$ Hz and
for a hydrodynamic gas, $\nu_H=\sqrt{10/3}\,\nu_\perp=2830(20)$
Hz.

The measured oscillation frequency of 3212(30) Hz for the
breathing mode of the noninteracting gas  is in excellent
agreement with the frequency 3200(20) Hz measured by the
parametric resonance method for the x-direction which is imaged in
the experiments. Hence, the parametric resonance method and the
breathing mode excitation method yield identical frequencies to
better than 1\%.

For the strongly interacting gas at 870 G and $T/T_F=0.17$, the
measured radial breathing mode frequency  $2837(05)$ Hz, is in
excellent agreement with the prediction $\nu_H=2830(20)$ for a
hydrodynamic Fermi gas, and it differs significantly from that of
the noninteracting gas and the weakly interacting Bose gas.
Similar results are obtained at 860 G and 880 G and 910 G at
slightly lower temperatures. For the axial direction, the measured
amplitude of the oscillation is consistent with zero.

On the molecular side, just below resonance at 815 G at
$T/T_F=0.14$, we obtain $\nu =2931(19)$, which is just above the
$\nu_H$, consistent with predictions that the response near
resonance is fermionic~\cite{Heiselberg2,Stringari}. At a much
lower field of 770 G, we find $\nu =3000(150)$ closer to the
predicted Bose frequency of 3100 (20) Hz. However, the data is not
of as high quality as that shown in Fig.~\ref{fig:2}.

Over the range of magnetic fields studied, our measured
oscillation frequencies $\nu(B)$ at the lowest temperatures show
the same magnetic field dependence as those of Rudi Grimm's
group~\cite{GrimmSF}. However, our data show a much smaller
frequency shift with respect to the expected hydrodynamic
frequency $\nu_H$.

We estimate the frequency shifts $\Delta\nu\equiv \nu
(\text{meas})-\nu(\text{actual})$ arising from anharmonicity in
the trapping potential~\cite{Stringarishift}. For the hydrodynamic
frequency, $\Delta\nu_H=
-(32/25)\sqrt{10/3}\,\nu_\perp\,M\omega_\perp^2x_{\text{rms}}^2/(b_H^2U)$,
where $U$ is the trap depth, $M$ is the $^6$Li mass, and
$\omega_\perp=2\pi\nu_\perp$. Here, $b_H$ is the hydrodynamic
expansion factor, 11.3 after 1 ms~\cite{expansion}. The shift in
the geometric mean of the transverse frequencies $\Delta\nu_\perp=
-(6/5)\nu_\perp\,M\omega_\perp^2x_{\text{rms}}^2/(b_B^2U)$, where
$b_B=10.3$ is the ballistic expansion factor at 1 ms.  Using
Table~\ref{table:1}, these results yield a net
$\Delta\nu\equiv\Delta\nu_H-\sqrt{10/3}\,\Delta\nu_\perp =+24$ Hz
at 870 G and $T/T_F=0.17$. For the three higher temperatures at
870 G, we find $\Delta\nu \simeq -35$ Hz at $T/T_F=0.33$ and 0.5,
and -16 Hz at $T/T_F=1.15$, consistent with the measured $\simeq
-60$ Hz shift below the lowest temperature data.

We have also investigated the effect of decreasing the oscillation
amplitude to less than 10\% by reducing $t_0$ to 25$\mu$s at 870 G
and $T/T_F=0.15$. For small amplitudes, the aspect ratio of the
cloud changes very little. Hence, we expect that the deformation
of the Fermi surface is very small, so that collisional behavior
is not induced. We obtain $\nu=2838(16)$ Hz, in precise agreement
with the results for $t_0=50\,\mu$s. The damping time is somewhat
increased to 6.00(1.4) ms, presumably due to the smaller energy
input, rather than reduced anharmonicity (since the frequency is
unchanged).

\begin{figure}
\begin{center}
\resizebox{3.5in}{!}{\includegraphics[0in,0.5 in][6.35
in,3.75in]{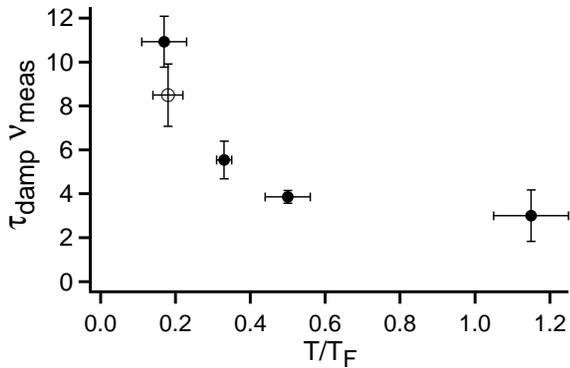}}
\end{center}
\caption{Product of the damping time and the measured breathing
mode frequency versus temperature.  The open circle shows the
result when the trap depth is increased by a factor of
4.1.}\label{fig:3}
\end{figure}

The measured damping time  shows a rapid increase with decreasing
temperature, Fig.~\ref{fig:3}. The open circle shows the result at
a trap depth increased by a factor of 4.1. The measured
hydrodynamic frequency scales as $\sqrt{4.1}$ within 3\% and the
product $\nu_{meas}\tau_{damp}$ is consistent with those at lower
trap depth. This is consistent with a  damping rate which scales
linearly with trap frequency, as expected for unitarity-limited
interactions, where the rate scales with the Fermi energy. Note
that anharmonicity cannot make a major contribution to the
temperature dependence of our damping rates: The anharmonic
contribution to $1/\tau_{damp}$ would be proportional to the shift
and therefore independent of trap depth for fixed $T/T_F$ and $N$.
Then, $\nu\tau_{damp}\propto\nu$, and the vertical position of the
open circle would increase by a factor of 2. Further, the
anharmonic contribution is identical for the three highest
temperature points where the mean cloud sizes are nearly the same.

We have attempted to model the data at 870 G without invoking
superfluidity. A first scenario is that the gas is nearly
collisionless at the lowest temperature, and the long damping time
and measured frequency are the result of collisionless mean field
evolution. A second  scenario is collisional hydrodynamics.

The collisionless mean field  scenario requires a large negative
mean field shift to explain the difference between the frequencies
of 3212(30) Hz and 2837(20) Hz measured for the noninteracting and
strongly interacting samples, respectively. However, for a
unitarity limited interparticle interaction  with a negative
$\beta=-0.55$~\cite{Carlson}, a Vlasov equation
model~\cite{Menotti} yields a  +90 Hz  shift relative to 3200 Hz,
while we observe a  -400 Hz  shift. Also, for our trap, the same
model shows that the coupling of the collisionless transverse
modes by the interaction would produce  a noticeable beat at 370
Hz with an amplitude minimum at 1 ms, which is not observed.
Hence, the data are inconsistent with the collisionless scenario.

To investigate the second scenario, we considered a collisional
hydrodynamic model, including two-body Pauli
blocking~\cite{Vichi,Unitarity}.  A small negative shift at the
higher temperatures might arise from the mean field in the
hydrodynamic limit~\cite{Heiselberg2,Stringari}, or from the
anharmonic shift described above. Neglecting these shifts, a
relaxation approximation model~\cite{Guery} can be used to
determine both the breathing mode frequency and the damping time
in terms of the measured trap oscillation frequencies, for an
arbitrary momentum relaxation rate. We find that a very large
momentum relaxation rate is needed to fit the 4 ms damping time of
the $T/T_F=0.17$ data in a collisionally hydrodynamic regime.
Then, the predicted damping time is large over a broad temperature
range, inconsistent with the observed rapid decrease in damping
time with temperature. Lowering the maximum relaxation rate, we
can fit the damping times at the two highest temperatures. In this
case, however, obtaining a 4 ms damping time requires a
temperature below $T/T_F=0.1$, i.e., a nearly collisionless
regime, inconsistent with observations as described above.

In conclusion, at our lowest temperatures, we observe a breathing
mode at precisely the hydrodynamic frequency as well as highly
anisotropic hydrodynamic expansion, as in our previous
experiments~\cite{OHaraScience,Menotti}. The damping time
increases rapidly as the temperature is lowered~\cite{expfit},
consistent with a transition from collisional to superfluid
hydrodynamics at a temperature between $0.2$ and $0.3$ $T_F$. On
the basis of the above arguments, we believe the data are not
consistent with either collisionless mean field evolution or
collisional hydrodynamics. It is therefore difficult to see how
the observations  can be explained without invoking superfluidity.

Recent theory describes the BCS-BEC crossover regime in terms of
very large fermionic pairs, comparable in size to the
interparticle
spacing~\cite{BruunCrossover,StoofCrossover,XiongCrossover}. Falco
and Stoof~\cite{StoofCrossover} predict  BEC-like or BCS-like
behavior {\em above} the Feshbach resonance, depending on whether
$\epsilon_b\equiv\hbar^2/(a^2\,m_{\text{atom}})$ is $\leq 2
k_BT_F$, or $\geq 2k_BT_F$~\cite{idea}. Near resonance, where
$\epsilon_b<<2k_BT_F$, the majority of fermionic pairs (either
Bose molecules or Cooper pairs) are very large. Hence, one expects
that the response of the system to compression is fermionic, and
scales with density as $n^{2/3}$~\cite{Ketterleidea}. This idea is
consistent with our measurements and with
predictions~\cite{Stringari,Heiselberg2}.

This research is supported by the  Chemical Sciences, Geosciences
and Biosciences Division of the Office of Basic Energy Sciences,
Office of Science, U. S. Department of Energy, the Physics
Divisions of the Army Research Office  and the National Science
Foundation, and  the Fundamental Physics in Microgravity Research
program of the National Aeronautics and Space Administration.


\begin{thebibliography}{99}
\bibitem{Houbiers}M.~Houbiers et al., {\it Phys. Rev. A\/} {\bf 57}, R1497
(1998).
\bibitem{Stwalley}W. C. Stwalley, {\it Phys. Rev. Lett.\/} {\bf 37}, 1628
(1976).
\bibitem{Tiesinga}B. E. Tsienga et al., {\it Phys. Rev. A\/} {\bf 47},
4114 (1993).

\bibitem{Heiselberg}H.~Heiselberg, {\it Phys. Rev. A\/} {\bf 63}, 043606
(2001).
\bibitem{MechStab}M.~E.~Gehm et al., {\it Phys. Rev. A\/} {\bf 68}, 011401
(2003).
\bibitem{Ho}T.-L. Ho, {\it Phys. Rev. Lett.\\/} {\bf 92}, 090402(2004).
\bibitem{Stoof1}Cooper pairing of $^6$Li atoms in the BCS approximation was
predicted by H. T. C. Stoof et al., {\it Phys. Rev. Lett.\/} {\bf
76}, 10 (1996); M. Houbiers et al., {\it Phys. Rev. A\/} {\bf 56},
4864 (1997).
\bibitem{Holland}M. Holland et al., {\it Phys. Rev. Lett.\/} {\bf 87}, 120406 (2001).
\bibitem{Timmermans}E. Timmermans et al., {\it Phys.  Lett.\/} A{\bf 285}, 228 (2001).
\bibitem{Griffin}Y. Ohashi and A. Griffin, {\it Phys. Rev. Lett.\/} {\bf 89}, 130402 (2002).
\bibitem{GrimmThomas}The first measurements of the breathing modes in a strongly
interacting Fermi gas were presented  at the Workshop on Fermi
Gases in Trento, Italy (March 4-6, 2004) by our group and by the
group of R. Grimm.
\bibitem{OHaraScience}K.~M. O'Hara et al., {\it Science\/} {\bf 298}, 2179 (2002).
\bibitem{Gupta}S. Gupta et al., {\it Phys. Rev. Lett.\/} {\bf 92}, 100401
(2004).
\bibitem{GreinerBEC}M. Greiner et al., {\it Nature\/} (London) {\bf 426},
537 (2003).
\bibitem{JochimBEC}S. Jochim et al., {\it Science\/} {\bf 302}, 2101
(2003).
\bibitem{ZwierleinBEC}M. Zwierlein et al., {\it Phys. Rev. Lett.\/}
{\bf 91}, 250401 (2003).
\bibitem{Salomon}T. Bourdel et al., cond-mat/0403091 (2004).
\bibitem{RegalCooper}C. A. Regal et al., {\it Phys. Rev. Lett.\/} {\bf
92}, 040403 (2004).
\bibitem{ZwierleinCooper}M. Zwierlein et al., cond-mat/0403049 (2004).
\bibitem{Vichi}L. Vichi, {\it J. Low Temp. Phys.\/} {\bf 121}, 177
(2000).
\bibitem{Heiselberg2}H. Heiselberg, cond-mat/0403041 (2004).
\bibitem{Stringari}S. Stringari, {\it Europhys. Lett.\/} {\bf 65}, 749
(2004).
\bibitem{zerocross}K. M. O'Hara et al., {\it Phys. Rev. A\/} {\bf 66},
041401 (2002).
\bibitem{Grimmzero}S. Jochim et al., {\it Phys. Rev.
Lett.\/} {\bf 89}, 273202 (2002).
\bibitem{GrimmSF}M. Bartenstein et al., cond-mat/0403716 (2004).
\bibitem{Stringarishift}This shift was derived for us by S. Stringari,
private communication.
\bibitem{expansion}Note that for the data at 19.2\% trap depth,
$b_H[0.8\,\text{ms}]=19.1$.
\bibitem{Carlson}J. Carlson et al., {\it Phys. Rev. Lett.\/} {\bf 91},
050401 (2003).
\bibitem{Menotti}C. Menotti et al., {\it Phys. Rev. Lett.\/} {\bf 89},
250402 (2002).
\bibitem{Unitarity}M. E. Gehm et al., {\it Phys. Rev. A\/} {\bf 68},
011603(R) (2003).
\bibitem{Guery}D. Guery-Odelin et al., {\it Phys. Rev. A\/} {\bf
60}, 4851 (1999).
\bibitem{expfit}The measured damping times at 870 G are well fit by
$\tau [\text{ms}]= 0.88 \,\exp [+0.25\,T_F/T]$.
\bibitem{BruunCrossover}G. M. Bruun, cond-mat/0401497 (2004).
\bibitem{StoofCrossover}G. M. Falco and H. T. C. Stoof,
cond-mat/0402579 (2004).
\bibitem{XiongCrossover}H. Xiong and S. Liu, cond-mat/0403336 (2004).
\bibitem{idea}We suspect that $\epsilon_b$  may play additional roles, i.e.,
compared to the  thermal energy, the trap energy level spacing,
trap laser recoil energy, etc.
\bibitem{Ketterleidea}W. Ketterle, private communication.

\end{thebibliography}
\end{document}